\documentclass{nature}
\usepackage{appendix}
\usepackage{threeparttable}
\usepackage{hyperref}
\usepackage{lineno}

\usepackage{amssymb, amsfonts, amsmath,graphicx,mathrsfs,xcolor}
\makeatletter
\let\saved@includegraphics\includegraphics
\AtBeginDocument{\let\includegraphics\saved@includegraphics}
\renewenvironment*{figure}{\@float{figure}}{\end@float}
\renewenvironment*{table}{\@float{table}}{\end@float}
\makeatother

\def\apj{Astrophys. J.}
\def\apjs{Astrophys. J. Suppl.}
\def\apjl{Astrophys. J. Lett.}
\def\aap{Astron. Astrophys.}
\def\mnras{Mon. Not. Roy. Astron. Soc.}
\def\prl{Phys. Rev. Lett.}
\def\prd{Phys. Rev. D}

\def \aj{The Astronomical Journal}
\def\pasa{Publ. Astron. Soc. Australia}

\begin{document}


\title{Ultrahigh-energy $\gamma$-ray emission  associated with the
tail of a bow-shock pulsar wind nebula} 

\date{\today}{}

\maketitle
Zhen Cao$^{1,2,3}$,
F. Aharonian$^{3,4,5,6}$,
Y.X. Bai$^{1,3}$,
Y.W. Bao$^{7}$,
D. Bastieri$^{8}$,
X.J. Bi$^{1,2,3}$,
Y.J. Bi$^{1,3}$,
W. Bian$^{7}$,
A.V. Bukevich$^{9}$,
C.M. Cai$^{10}$
% ...
(The LHAASO Collaboration)

$^{1}$ Key Laboratory of Particle Astrophysics \& Experimental Physics Division \& Computing Center, Institute of High Energy Physics, Chinese Academy of Sciences, 100049 Beijing, China\\
$^{2}$ University of Chinese Academy of Sciences, 100049 Beijing, China\\
$^{3}$ TIANFU Cosmic Ray Research Center, Chengdu, Sichuan,  China\\
$^{4}$ University of Science and Technology of China, 230026 Hefei, Anhui, China\\
$^{5}$ Yerevan State University, 1 Alek Manukyan Street, Yerevan 0025, Armeni a\\
$^{6}$ Max-Planck-Institut for Nuclear Physics, P.O. Box 103980, 69029  Heidelberg, Germany\\
$^{7}$ Tsung-Dao Lee Institute \& School of Physics and Astronomy, Shanghai Jiao Tong University, 200240 Shanghai, China\\
$^{8}$ Center for Astrophysics, Guangzhou University, 510006 Guangzhou, Guangdong, China\\
$^{9}$ Institute for Nuclear Research of Russian Academy of Sciences, 117312 Moscow, Russia\\
$^{10}$ School of Physical Science and Technology \&  School of Information Science and Technology, Southwest Jiaotong University, 610031 Chengdu, Sichuan, China\\
% ...


\begin{abstract}

In this study, We present a comprehensive analysis of an unidentified point-like ultra-high-energy (UHE) gamma-ray source, designated as 1LHAASO~J1740+0948u, situated in the vicinity of the middle-aged pulsar PSR~J1740+1000. The detection significance reaches $17.1\sigma$ ($9.4\sigma$) above $25\,\rm{TeV}$ ($100\,\rm{TeV}$).The energy spectrum of the source extends up to $300\,\rm{TeV}$ and is well-described by a log-parabola function with $N_0=(1.93\pm0.23)\times10^{-16}\,\rm{TeV}^{-1}\,\rm{cm}^{-2}\,\rm{s^{-1}}$, $\alpha=2.14\pm0.27$, and $\beta=1.20\pm0.41$ at $E_0=30\,\rm{TeV}$. The associated pulsar, PSR~J1740+1000, resides at a high Galactic latitude and powers a bow-shock pulsar wind nebula (BSPWN) with an extended X-ray tail. 
The best-fit position of the gamma-ray source appears to be shifted by 0.2$^{\circ}$ with respect to the pulsar position. As the (i) currently identified pulsar halos do not demonstrate such offsets, and (ii) centroid of the gamma-ray emission is approximately located at the extension of the X-ray tail, we speculate that the UHE $\gamma$-ray emission may originate from re-accelerated electron/positron pairs that are advected away in the bow-shock tail. 
\end{abstract}

\section{\bf INTRODUCTION}
Pulsars are highly magnetized neutron stars with fast spin. They launch relativistic pulsar winds composed of electron/positron pairs at the expense of consuming the rotational energy of the pulsars. The interaction of the powerful pulsar wind with surrounding medium abruptly decelerates the wind at a termination shock (TS), where pairs in the wind can be further accelerated and form the pulsar wind nebulae (PWNe) downstream. PWNe are believed to be efficient sites of particle acceleration, as evidenced by the detection of nonthermal X-ray and $\gamma$-ray emission from many PWNe, which originates from energetic pairs through synchrotron radiation and inverse Compton scattering, respectively. Particularly, PWNe driven by young energetic pulsars, such as the Crab Nebula, can even accelerate pairs to energies at the PeV  scale\cite{ref_24}. Efficient particle acceleration in PWNe can persist for extended periods, potentially spanning millions of years. The recent discovery of extended TeV $\gamma$-ray emissions around some middle-aged pulsars\cite{ref_1,ref_2}, referred to as “TeV halos (or pulsar halo)”, suggest that acceleration of $\sim 100 \,\rm{TeV}$ pairs is still possible in PWNe older than $100\,\rm{kyr}$. 

Note that pulsars may receive natal kicks in their progenitor supernova explosions, with typical velocities of several hundred $\rm{km \, s^{-1}}$, as shown in the catalog of 233 pulsars \cite{ref_3}. Thus, at the middle-age stage, these pulsars may have left the associated supernova remnants and traverse in the interstellar medium(ISM). In general, their proper speeds are much faster than the sound speed in the ISM, which is typically a few tens of $\rm{km \, s^{-1}}$. As a result, the proper motion drives a bow shock ahead in the direction of their movement, and the ram pressure exerted on the PWN by the incoming ISM conﬁnes the PWN to the direction opposite to its proper motion, forming a “tail” trailing behind, as often observed in the X-ray and radio bands. 

Observations of X-ray emission from the bow-shock tail indicate that accelerated high-energy pairs are transported away from the PWN. In principle, we may expect TeV $\gamma$-ray emission from the bow-shock tail as well, emitted by the same population of escaping pairs through upscattering the cosmic microwave background (CMB) and the interstellar infrared background (IRB), especially considering that the magnetic field strength may drop at the long tail, where is far away from the PWN. The emission of these energetic pairs not only serves as an informative diagnostic of pulsar wind properties but also provides a clue on how these energetic particles escape the PWNe, which is important for understand the generation of TeV Halos. It would also provide insights into the particle transport in Vela X PWN from the compact PWN to distant regions\cite{vela_2006,vela_2012}. However, $\gamma$-ray emission has not been clearly detected from the long tail of BSPWN so far \cite{ref_4}.

Herein, we conducted a search for gamma-ray photons in the vicinity of PSR~J1740+1000, which exhibits a significant tail due to its supersonic motion in the ISM. The pulsar is located at a high Galactic latitude of over $20^{\circ}$, and its proper motion direction slightly points to the Galactic plane, indicating that it might have formed in the Galactic halo rather than moving from the Galactic plane \cite{ref_5,ref_8}. This presents a rare opportunity for us to investigate UHE $\gamma$-ray emission from a high Galactic latitude environment and a pulsar with a halo star progenitor. PSR~J1740+1000 is a middle-aged pulsar with a characteristic age of about $114\,\rm{kyr}$, located at a distance of $1.4\,\rm{kpc}$ from Earth according to the dispersion measurement. Its current rotational period is about $0.154\,\rm{s}$, and the spin-down luminosity is $2.32 \times 10^{35} \, \rm{erg \, s^{-1}}$. The pulsar was initially detected in the radio band during an Arecibo survey, but it traversed through the North Polar Spur (NPS), which represents the brightest region of the expansive radio feature known as Loop I. Consequently, there is a lack of deep radio observation of this source.\cite{ref_5}

This X-ray PWN shows a tail extending southwest approximately $6'$ from the pulsar's position\cite{ref_8,ref_4}. Initially, the tail appears conical with an opening angle of 15${^{\circ}}$, and at a distance greater than about $3'$ it appears to taper off and maintain a cylindrical shape. The X-ray spectrum can be fitted by an absorbed power-law with an index $\Gamma = 1.75 \pm 0.04$ and a normalization $N_{0} = (3.41 \pm 0.09) \times 10^{-5} \,\rm{keV^{-1} cm^{-2} s^{-1}}$ at $1 \,\rm{keV}$, corresponding to a flux of $F_{X} = (1.93 \pm 0.06) \times 10^{-13} \,\rm{erg \, cm^{-2} s^{-1}}$ in $0.3-10 \,\rm{keV}$.
In the direction of pulsar's proper motion, there is a GeV source named 4FGL~J1740.5+1005, identified as the $\gamma$-ray emission from the pulsar\footnote{see \href{https://confluence.slac.stanford.edu/display/GLAMCOG/Public+List+of+LAT-Detected+Gamma-Ray+Pulsars}{Public List of LAT-Detected Gamma-Ray Pulsars}}, since a pulsed component has been detected with an H-test significance of 31$\sigma$\cite{ref_9}. 

At the TeV $\gamma$-ray band, The VERITAS collaboration observed the region around PSR~J1740+1000 with a total live time of  12.8\,hr, but reported only an upper limit of a few times $10^{-13}\,\rm cm^{-2}s^{-1}$ in the energy band between 1 and 10 TeV. HAWC detected an unresolved $\gamma$-ray source about $0.1^{\circ}$ southwest of the pulsar's position, named 3HWC J1739+099 \cite{ref_12}. The spectrum is described by a power-law function with an index of $1.98_{-0.16}^{+0.25}$, and the flux is found to be $3.3_{-1.4}^{+2.2}\times 10^{-15}\,\rm{TeV}^{-1}\,\rm{cm}^{-2}\,\rm{s}^{-1}$ at $7\,\rm{TeV}$. The source is considered a candidate for a pulsar halo, although the relatively low significance of the emission hinders further identification of the source's origin\cite{ref_12}. 
 
The exceptional sensitivity of the Large High Altitude Air Shower Observatory (LHAASO) at high energies, coupled with its large field of view, makes it a valuable tool for detecting and characterizing emissions from the extended tail of the BSPWN. Details of the LHAASO observatory can be found in the \nameref{sec:sup_info}. In LHAASO's first catalog\cite{ref_16}, a point-like UHE $\gamma$-ray source, 1LHAASO~J1740+0948u, is discovered with a spatial offset of $0.21^\circ$ from the position of the pulsar, based on approximately 3 years of live time(1/2, 3/4 and full array). The spectrum extends beyond 100\,TeV without a known counterpart at other wavelengths. In the remainder of this paper, we will perform a dedicated analysis of LHAASO\rq s  data and explore the nature of the source. 

\section{\bf Method} 
In this work, we used over 1200 days of data from LHAASO-KM2A and over 900 days of data from LHAASO-WCDA. Adhering to the LHAASO analysis pipeline, the data preprocessing procedures, including event selection, reconstruction, and $\gamma$-proton discrimination, followed the methodologies outlined in previous performance evaluation publications by LHAASO \cite{ref_13,2024arXiv240511826H}. In gamma astronomy, other hadronic components serve as background. Taking advantage of the exceptional capabilities of the muon detectors, we achieved a hadron survival rate of only 0.0001\% at 150 TeV, meaning almost background-free data at high energies. Subsequently, we conducted an analysis using the photon data selected post $\gamma$-proton discrimination. For further information regarding LHAASO and its dataset, please refer to Section \nameref{sec:sup_info}.

We estimate the background level ($N_{off}$) using the direct integration method \cite{ref_14}. This method involves counting the events that originate from the same spatial direction as the source, but during time intervals when the target source is not in that direction. To achieve the best performance, we only use the events with zenith angle below $50^{\circ}$. Since the $\gamma$-ray source located at a high galactic latitude($>20^{\circ}$), we ignore the diffuse $\gamma$-ray emission here. We then use the 3D likelihood method to fit both the spatial distribution and the spectral energy distribution (SED). The significance of the source was evaluated using the test statistic (TS), defined as twice the logarithmic likelihood ratio, i.e., $\rm{TS}=2\rm{ln}(\mathcal{L}_{s+b}/\mathcal{L}_b)$, where $\mathcal{L}_{\rm{s+b}}$ represents the maximum likelihood for the source signal plus background hypothesis, and $\mathcal{L}_b$ represents the null hypothesis. The TS value follows a chi-squared distribution, with the number of degrees of freedom equal to the difference in the number of free parameters between the hypotheses\cite{ref_15}. For a point source with a fixed position, which has only one free parameter (i.e.,  the normalization, assuming a fixed spectral distribution), the pretrial significance is $\sqrt{\rm{TS}}\sigma$. The value in each pixel of the TS map is calculated using this method.

In this study, we test several commonly used morphology templates, such as point source, 2D Gaussian and disk. The point source model assumes that only the detector's Point-Spread Function (PSF, commonly represented as a Gaussian distribution) is considered, whereas other extended templates require convolving the specific spatial distribution with the PSF. We fit the spectrum using the forward-folding method, where the expected number of photons in each energy bin is based on the simulation of the instrument response for incident events\cite{2024arXiv240404801C}. We only provide spectral data points for energy bins where the TS value is greater than 4, otherwise we assign upper-limit values. The optimal fitting result occurs when the appropriate spatial distribution and spectrum shape yield the maximum significance. 

\section{\bf Result}\label{sec:result}
\begin{figure}[h]
	\begin{center}
        \includegraphics[width=0.95\linewidth]{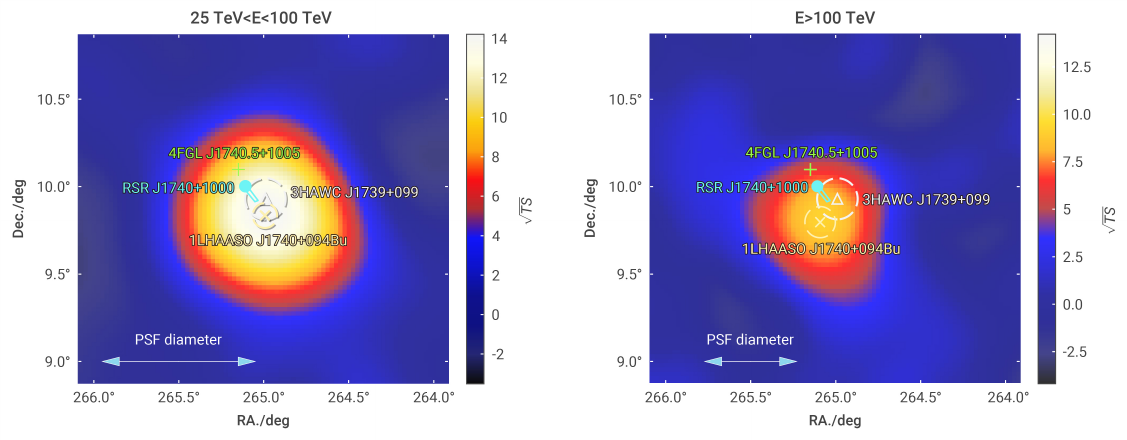}
		\caption{The significance map around 1LHAASO~J1740+0948u with $25\,\rm{TeV}<E<100\,\rm{TeV}$  (right: $> 100\,\rm{TeV}$) is shown in the figure. The green cross symbol represents the best-fit position of the $\gamma$-ray source in point template, Meanwhile, the grey triangle represents the position provided by the 3HAWC catalog. The cyan solid circle symbol corresponds to the position of PSR~J1740+1000, and the lime plus symbol represents the location of the Fermi-LAT source. The green and grey dashed line circle represents the $95\%$ error bounds of the LHAASO and HAWC source. Notably, the cyan striped shape extending from the pulsar indicates a $6^{'}$ X-ray tail. The double directional arrow is the 68\% PSF diameter.}
		\label{fig:TSmap}
	\end{center}
\end{figure}

To avoid potential influence from large bubbles or other sources in the surrounding region, we have selected a large region of interest (ROI) as an $8.4^\circ \times 8^\circ$ as a rectangle for further analysis. Within this extended area, only a bright, isolated gamma-ray source, 1LHAASO~J1740+0948u, was observed near PSR~J1740+1000. This gamma-ray source is rare in the first LHAASO catalog because it is located at such a high Galactic latitude (Gb $>20^{\circ}$). 1LHAASO J1740+0948u is primarily detected at higher energies ($>$25 TeV) by LHAASO-KM2A, while at lower energies (a few TeV), LHAASO-WCDA does not show significant excess signal. Here, we present two TS maps of LHAASO-KM2A in different energy intervals centered on the source as shown in Fig.~\ref{fig:TSmap}.


Various morphology templates were tested for 1LHAASO~J1740+0948u using the KM2A data. Table. \ref{tab:morphology} in the Supplementary Information section lists the best-fit parameters for these templates. Despite the increased degree of freedom, these models did not significantly improve upon the point-like source case. Consequently, we set the $95\%$ confidence level upper limit of the source extension using the methodology outlined by Cao et al.\cite{ref_16}, yielding a conservative value of approximately $0.147^\circ$.  Using the point template, the optimal position is determined to be $\rm{(R.A., Decl)}=(265.03^{\circ} \pm 0.02^{\circ}, 9.82^{\circ} \pm 0.02^{\circ})$, with a significance of 17.1$\sigma$ above $25\,\rm{TeV}$. Notably, this significance remains high at 9.4$\sigma$ over 100TeV, corresponding to a post-trial significance of 16.7(8.6)$\sigma$ over 25(100) TeV. As shown in Fig.\ref{fig:TSmap}, our fitted position is consistent with the HAWC observation's position within the $95\%$ position error circle.
The best-fit position of the source at $E > 100$ TeV and that in $25 {\rm TeV}-100\,\rm{TeV}$ shows a slight deviation of approximately $0.1^{\circ}$, which is within the acceptable $95\%$ error circle.
\begin{table}
\caption{Morphological models tested for the 1LHAASO~J1740+0948u above 25 TeV.}
\begin{center}
     \begin{tabular}{lccccccc}
        \hline
        Template & R.A & Dec & $\rm{Extension}^{\rm{a}}$ & $\rm TS_{ext}^{\rm b}$ & $\rm{N_{p}}^{c}$ \\ \hline
       Point source           & $265.03 \pm 0.02$  & $9.82 \pm 0.02$ & -  & - & 5 \\ 
        Gaussian           & $265.03 \pm 0.02$  & $9.81 \pm 0.02$ & $0.09 \pm 0.04$    & 1.03 & 6 \\ 
         Disk    & $265.03 \pm 0.02$ & $9.81 \pm 0.02$ &  $0.17 \pm 0.03$   & 1.09 &  6 \\ 
         Diffuse & $265.03 \pm 0.02$ & $9.81 \pm 0.02$ & $0.18 \pm 0.09$ & 0.96 & 6 \\  
         \hline
    \end{tabular}
    \end{center}
    \begin{tablenotes}
    \footnotesize
    \item $^a$ Extension in the disk and Gaussian models represents the radius containing 68\% of the flux of the tested models.
     \item $^b$ ${\rm TS}_{\rm{ext}}=2\rm{ln}(\mathcal{L}_{\rm{ext}}/\mathcal{L}_{\rm{ps}})$ is defined where $\mathcal{L}_{\rm{ext}}$ is the maximum likelihood value for the extend model and $\mathcal{L}_{\rm{ps}}$ is the maximum likelihood value for the point-like model. According to Ref.\cite{ref_23}, the criterion to deﬁne a source as being extended is $TS_{\rm{ext}} \geq 16$.
    \item $^c$ $N_{\rm p}$ represents the number of degrees of freedom for each model.
    \end{tablenotes}
    \label{tab:morphology}
\end{table}

\begin{figure}[!htpb]
    \centering
    \includegraphics[width=0.85\textwidth]{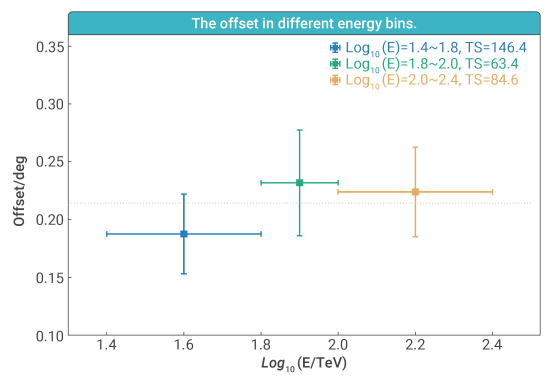}
    \caption{Offsets between PSR J1740+1000 and the LHAASO source for an energy range of log10(E/TeV) = 1.4–1.8 (cyan), 1.8–2.0 (lime), and 2.0–2.4 (yellow).}
    \label{fig:offset with energy}
\end{figure}

Notably, we found that PSR~J1740+1000 is significantly offset from the best-fit position of the KM2A source, lying outside the 95\% position error circle, even at energies exceeding 100 TeV. The fit results indicate an offset of $\Delta = 0.219^{\circ} \pm 0.025^{\circ}{\rm{stat}} \pm 0.018^{\circ}{\rm{sys}}$ above 25 TeV, with the systematic error accounting for the deviation from the Crab position (see Fig. \ref{fig:err} in the Supplementary Information section). To analyze the energy-dependent behavior of the offset, we selected three significant energy bins: $\log_{10}(E/\rm{TeV})=1.4-1.8, 1.8-2.0, 2.0-2.4$, to investigate the offset, with results are shown in Fig. \ref{fig:offset with energy}. Results show that the large offset indeed exists in all energy bins, and position offsets from the pulsar are consistent with each other. No significant energy-dependent evolution of the offset has been observed with the limited significance. 

\begin{figure}[h]
	\begin{center}
		\includegraphics[width=0.85\textwidth]{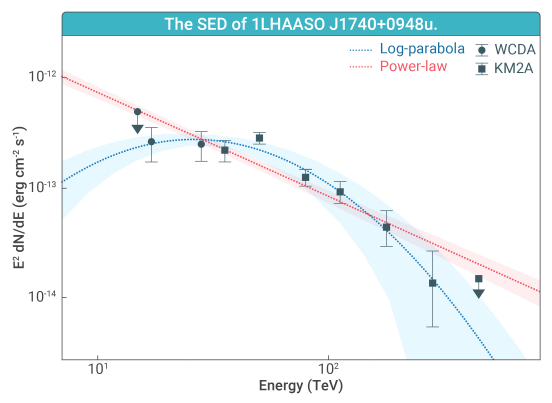} 
		\caption{Spectral data of 1LHAASO J1740+0948u from LHAASO-WCDA (black circles) and LHAASO-KM2A (black squares) fitted with a single power law (red-dotted curve) and a log-parabola function (blue-dotted-line). The red and blue shaded bands show the 1$\sigma$ error range.}
		\label{fig:SED_logparabola}
	\end{center}
\end{figure}

We further extracted the spectrum of 1LHAASO J1740+0948u(Fig. \ref{fig:SED_logparabola}). Although LHAASO-WCDA did not detect a significant signal, we can still assume a point source based on the best-fit position from LHAASO-KM2A to calculate spectral points or upper limits for LHAASO-WCDA (see Methods section). Assuming a power-law function, the best-fit spectrum, combined both LHAASO-WCDA and LHAASO-KM2A data show below: 

\begin{equation}
    dN/dE=(1.86\pm0.17)\times10^{-16}(E/30\ \rm{TeV})^{-2.96\pm0.07}\,\rm{TeV^{-1}\,cm^{-2}\,s^{-1}}
\end{equation}
We also attempted to fit the spectrum with a log-parabola function $dN/dE=N_0(E/E_0)^{-\alpha-\beta\rm{log}_{10}(E/E_0)}$ with $E_0=30{\rm\ TeV}$. The obtained best-fit parameters were $N_0=(1.93\pm0.23)\times10^{-16}\,\rm{TeV^{-1}\,cm^{-2}\,s^{-1}}$, $\alpha=2.14\pm0.27$, and $\beta=1.20\pm0.41$. This resulted in an improvement of approximately $3.5\sigma$ compared to a single power-law fit. The log-parabola spectrum indicates a gradually steepening spectrum above $30\,\rm{TeV}$, but the spectrum still extends up to approximately 300 TeV, indicating the operation of an efficient acceleration process in the source.

To investigate whether the offset is due to a systematic effect introduced by updates to the LHAASO-KM2A array during construction, we extracted the 1/2, 3/4, and full array data separately for the same analysis. The results of the 
offset are shown in \nameref{sec:sup_info} Fig.\ref{fig:diffarr}. We observe that the flux and the position are generally consistent across the different array stages. In particular, the position offset remains stable across different construction stages of the LHAASO array.

\begin{figure}[!htpb]
	\begin{center}
		\includegraphics[width=0.85\textwidth]{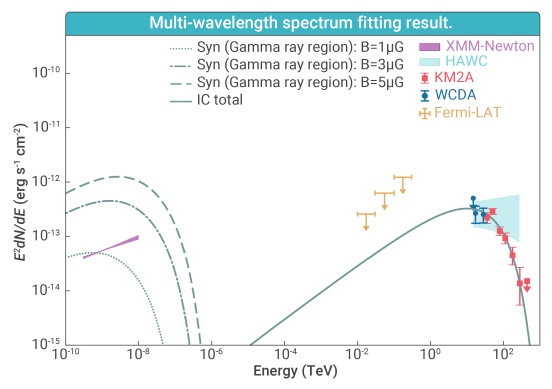}
		\caption{Phenomenological fitting to the spectrum of 1LHAASO~J1740+0948u.  Blue circles and red squares represent the LHAASO-WCDA and LHAASO-KM2A data, respectively.  Orange arrows indicate the Fermi-LAT flux upper limits of the source (please refer to \nameref{sec:sup_info} for detailed information regarding Fermi-LAT's data analysis process). For reference, we also show the flux of 3HWC J1739+099 measured by HAWC with the grey shaded area, and the X-ray flux of the PWN measured by XMM-Newton, noting that latter is not from the same region of 1LHAASO~J1740+0948u. The black solid curve shows the IC emission from a population of electrons with a total energy of $2.6\times 10^{45}$\,erg, with spectral distribution of $dN/dE \propto E^{-2}\exp(-E/{\rm 110\,TeV})$. The dotted, dash-dot, and dash lines represent the expected synchrotron radiation of the same population of electrons in the magnetic field of 1, 3, and 5\,$\rm{\mu G}$ respectively.}
		\label{fig:SED}
	\end{center}
\end{figure}

We interpret the observed $\gamma$-ray emission as the IC emission of electrons and perform a phenomenological fit to the gamma-ray spectrum using the Python package NAIMA\cite{ref_16_1,ref_16_2,ref_16_3}, as shown in Fig.~\ref{fig:SED}. The electron spectrum used follows a power-law distribution with a high-energy cutoff, defined as $dN/dE = N_eE^{-2}\exp(-E/E_c)$, with a minimum energy of 0.1 GeV. Here, $N_e$ is the normalization factor, and $E_c$ is the cutoff energy. 

In this analysis, we incorporate three seed photon fields: the cosmic microwave background (CMB), infrared, and optical, with temperatures $\rm{T}=[2.73,30,5000]$ K and energy densities $\rm{U}=[0.26,0.25,0.5]\,\rm{eV\,cm^{-3}}$ based on the interstellar radiation field (ISRF) model\cite{2022MNRAS.509.2339N}. We find $N_e=1.2\times 10^{30}\,\rm{eV^{-1}}$, resulting in a total electron energy of $2.6\times 10^{45}$\, erg, which is reasonable for a source powered by the rotational energy of PSR~J1740+1000. The cutoff energy $E_c$ is found to be 110 TeV, suggesting an effective particle acceleration mechanism. 

Given the distinct position of the KM2A source relative to the PWN of PSR~J1740+1000, the X-ray emission observed from the PWN does not arise from the same position as the gamma-ray emission detected by LHAASO. However, there should be an X-ray counterpart of the LHAASO source generated by the synchrotron radiation from the same electrons responsible for the TeV emission. The X-ray flux depends on the magnetic field strength $B$ in the source. We illustrate the anticipated X-ray flux for $B=1,\, 3,\, 5\,\mu$G in Fig.~\ref{fig:SED}. Additionally, we show the X-ray flux of the tail for reference. It is evident that the expected X-ray flux from the LHAASO source exceeds that of the PWN for $B=3\,\mu$G and $5\,\mu$G, indicating it could be detectable with an appropriate exposure time by current X-ray satellites such as XMM-Newton and Chandra.

\section{\bf DISCUSSION}
The best-fit position of 1LHAASO~J1740+0948u deviates significantly from PSR~J1740+1000. However, aside from the pulsar, we do not find any other potential astrophysical counterparts at other wavelengths within 1.5$^{\circ}$ from the $\gamma$-ray source. Notably, the presence of UHE photons up to $300\,\rm{TeV}$ from this source rules out the possibility of an extragalactic origin. We estimate the density of pulsars with a spin-down luminosity equal to or greater than PSR~J1740+1000 in a $20^{\circ}\times5^{\circ}$ region centered on the source according to the ATNF catalog\cite{ANTF}. Considering the position is located near the tail and within an angle of about $60^{\circ}$, multiplying with the possibility of the tail region location, the overall chance pulsar coincidentally appearing within $0.2^\circ$ of the source is estimated to be $0.07\%$(if a cut on $\dot{\rm{E}}/d^2> 10^{34}\, \rm{erg\,s^{-1}\,pc^{-2}}$ is applied), following the method employed in Ref\cite{1998ASPC..144...39M,ref_18}. A brief overview of the calculation process for this probability can be found in \nameref{sec:sup_info}. The low probability suggests a likely physical association between PSR~J1740+1000 and 1LHAASO~J1740+0948u. Thus, there must be a complex mechanism beyond our current understanding responsible for the spatial offset.

\noindent {\bf Hadronic Origin?}\\
One possible explanation for the offset is a molecular cloud illuminated by protons accelerated in the PWN, which then propagate there. Such an explanation require the presence of a matter target at the position of the source. However, the pulsar is located at a high Galactic latitude, approximately 400 parsecs away from the Galactic plane, where the column density of matter is expected to be only $8\times10^{20} \,\rm {cm^{-2}}$ at the source\rq s position while fitting in the X-ray tail\cite{ref_4}. Most CO surveys do not cover such high latitudes. Planck dust data only offer observations along the line of sight\cite{plank_dust1,plank_dust2}, but as shown in Fig.\ref{fig:dust} from \nameref{sec:sup_info}, the column density is consistent with X-ray fitting result around $10^{21} \,\rm {cm^{-2}}$. There appears to be no substantial matter located at the position of the $\gamma$-ray source. This sparse distribution of material is inadequate to account for such a high $\gamma$-ray luminosity. Therefore, the likelihood of the hadronic scenario in this context is low. Furthermore, how many protons can be accelerated to high energies by pulsars remains an open question. To conclusively rule out the hadronic origin, precise observational data of CO at the distance of the source is essential.

\noindent {\bf Pulsar halo scenario}\\
PSR~J1740+1000 is similar to other pulsars powering halos  such as Geminga, Monogem, and PSR J0622+3749\cite{ref_1,ref_2} in terms of characteristic age and rotation period. The observed X-ray BSPWN indeed suggests the acceleration of high-energy electrons in the PWN. However, two issues prevent an unambiguous identification of the source as a pulsar halo:

(i) TeV halos are typically associated with extended $\gamma$-ray emission observed around middle-aged pulsars. Previously measured pulsar halos or candidates, such as those around Geminga, Monogem, and PSR~J0622+3749
all exhibit pronounced extended structures. For 1LHAASO~J1740+0948u, as a point-like source, we may obtain an upper limit of $0.147^{\circ}$ for the extension, corresponding to about 3.6\,pc given the pulsar's distance of 1.4\,kpc. This is much smaller than the typical size of a pulsar halo, usually several tens of parsecs. Explaining such a compact size under the pulsar halo scenario would require a very short diffusion length of electrons, either due to slow diffusion or rapid cooling of electrons. In the slow diffusion scenario, a diffusion coefficient smaller than the Bohm limit would be required, which could be attributed to the cross-field diffusion with the average magnetic field direction approximately aligned with our line of sight toward the pulsar \cite{Liu19}. Typically, particle propagation in pulsar halos follows a diffusion model. We use a fitting form of the morphological distribution from a diffusion model under the approximation of continuous injection from a point source, as described in the work by LHAASO collaboration (2021)\cite{ref_2},

\begin{equation}
    f(\theta) \propto \frac{1}{\theta_{d}(\theta + 0.085\theta_{d})}\rm{exp}[-1.54(\theta/\theta_{d})^{1.52}],
\end{equation} 
to fit the LHAASO observed morphology. Here $\theta$ is the angular distance from the source position, $\theta_{d} = 180^{\circ}/\pi \cdot 2\sqrt{D(E_{e})t_{E}}/d$ is the typical diffusion extension where $D(E_{e})$ is diffusion coefficient. Although, the $\theta_d$ is an energy-dependent parameter, due to limited statistics, we treat it as constant, which should represent the typical value for the considered energy range(i.e., above 25 TeV). The test results are presented in Table \ref{tab:morphology}, showing no significant improvement of the diffusion model compared to the point source model.

(ii) The spatial offset between the source's position and the pulsar's position cannot be well explained with the present theoretical models of pulsar halos. In principle, the fast proper motion of pulsar could introduce asymmetric morphology of the pulsar halo \cite{ref_20, Johan19, Dimauro19}. The precise measurement of PSR~J1740+1000's velocity has not been achieved thus far. An estimation of the transverse velocity based on the interstellar scintillation\cite{ref_5} suggested $127 \, \rm{km\,s^{-1}} < v_{\rm{T}} < 240 \, \rm{km\,s^{-1}}$. An upper limit of proper motion velocity was reported as 60 mas/yr (341.3\,km/s at 1.2\,kpc), using 10 years of data from Chandra \cite{ref_21}. The resulting position offset was determined to be $\Delta = 0.219^{\circ} \pm 0.025^{\circ}_{\rm{stat}} \pm 0.018^{\circ}_{\rm{sys}}$ above 25\,TeV(Fig.\ref{fig:offset with energy} and Fig.\ref{fig:err} in \nameref{sec:sup_info}). If the offset were caused by the proper motion, the maximum offset could occur if the pulsar impulsively injects a group of electrons and move further away by a time equal to the cooling time of injected electrons, resulting in a rough offset of ${\rm arctan}(v_{\rm{T}} T_{\rm{cool}}/d)$ \cite{ref_19}. It would require over 500\,km/s for the transverse proper motion velocity of the pulsar given the typical interstellar magnetic field strength of $3 \,\mu\rm{G}$. Considering a continuous electron injection  scenario as \cite{ref_20}, it would require an even higher velocity. A recent study of the so-called ``mirage'' sources \cite{Bao24a, Bao24b} might provide a possible interpretation of the compact size and the spatial offset of PSR~J1740+1000. By simulating asymmetric diffusion of escaping electrons over multiple magnetic field coherence, it is found that the emission of these electrons may form a source with a significant offset from its accelerator due to the visual projection effect, given an applicable geometry of the local magnetic field. 

\noindent {\bf Particle re-acceleration at the bow-shock tail}\\
The difficulties in explaining 1LHAASO~J1740+0948u with both the hadronic model and the halo model have refocused our attention on the bow-shock tail. Although the measured X-ray tail length is limited to about 0.1 degrees (2.4 pc) due to the restricted field of view, and a positional offset still exists with respect to the LHAASO source, it remains possible for electrons to propagate to more distant regions. One possible indication of this is that the extension direction of the tail points toward the location of the LHAASO source. Furthermore, longer bow-shock tails have been observed in other bow-shock PWNe, such as PSR~J0002+6216 and J1638-4713 \cite{ref_22,2024PASA...41...32L}, where radio observations have inferred tail lengths of approximately 7 pc and 21 pc, respectively. On the other hand, the lack of observed $\gamma$-ray emission from the currently detected bow-shock tail suggests that the IC emission efficiency of electrons/positrons in the bow-shock tail may not be high, possibly due to suppression by a relatively strong magnetic field. However, MHD simulations\cite{Barkov19} indicate that in the process of magnetic reconnection, the downstream flow undergoes repeated compressions, which may lead to the formation of additional shocks, allowing particles to be re-accelerated within the tail flow and potentially produce stronger IC emission. Additionally, a lower magnetic field may exist in the outer regions, facilitating IC emission. Nevertheless, it remains challenging to provide a quantitative description of the offset between the tail and TeV emission, which requires further research, such as deep X-ray observations of this region, to better constrain the high-energy processes of the BSPWN.
The properties of PWNe are influenced by factors such as supernova energy and the characteristics of the surrounding medium. In particular, for PWNe embedded in supernova remnants, it is often challenging to distinguish these different components. In BSPWNe, however, the pulsar usually has a high kick velocity and, after a period of time, escapes from the supernova remnant and interacts with the ISM. This means that BSPWNe have relatively simple geometrical structures, where there is a clear separation of different components created by the newly shocked winds and the synchrotron flow carried downstream by the ram pressure. This makes them ideal systems for studying particle acceleration, morphology, and the evolution of pulsar winds. Synchrotron radiation in the tail of BSPWNe has already been detected in X-rays, and detecting IC radiation from the tail would provide electrons/positrons distribution information along the tail. This is crucial for understanding particle acceleration, transport, and related processes in PWNe. In our case, the source shows a similar physical scale to that of the nebula's tail and exhibits a collimated structure aligned with it. Our observations facilitate comprehensive detection along the elongated tail, addressing both spatial and energetic dimensions, suggesting the presence of effective re-acceleration sites for electrons and positrons within the tail. This mechanism may also be applicable to other bow shock systems, including Vela-like objects and even stellar bow shocks\cite{stellar_bow_shocks}. Moreover, this implies that BSPWN play a crucial role in the production of electrons and positrons, which can escape and undergo re-acceleration, potentially providing insights into the positron excess puzzle.

\section{Conclusions}

In summary, a UHE $\gamma$-ray source 1LHAASO~J1740+0948u has been detected in the vicinity of the high-latitude pulsar J1740+1000, with approximately three years of data of LHAASO-KM2A. The source exhibits a significance of $17.1 \sigma$ above $25 \rm{TeV}$ and $9.4 \sigma$ above $100 \rm{TeV}$. The pulsar appears be the most promising astrophysical object powering this UHE source. However, the point-like nature of 1LHAASO~J1740+0948u and the $0.2^\circ$ offset between the pulsar and the source make it challenging to provide a clear physical picture. We suggested that re-acceleration of electrons/positrons advected to the downstream of the bow-shock tail flow might provide an explanation, but the scenario need to be elucidated with detailed studies in the future. A pulsar halo might also work if considering asymmetric propagation of escaping electrons in the surrounding medium. On the other hand, multi-wavelength observations of the source will be crucial to unravel the puzzle. A search for correlated molecular or atomic gas content can be used to determine the radiation mechanism. Morphological and spectral studies by X-ray observations and imaging air Cherenkov telescope with high angular resolution will help to clarify the origin of the UHE emission and probe the physics inside the source.

\noindent {\bf Acknowledgements}\\
We would like to thank all staff members who work at the LHAASO site above 4400 meter above the sea level year round to maintain the detector and keep the water recycling system, electricity power supply and other components of the experiment operating smoothly. We are grateful to Chengdu Management Committee of Tianfu New Area for the constant financial support for research with LHAASO data. We appreciate the computing and data service support provided by the National High Energy Physics Data Center for the data analysis in this paper. This research work is supported by the following grants: The National Natural Science Foundation of China No.12393851, No.12393852, No.12393853, No.12393854, No.12205314, No.12105301, No.12305120, No.12261160362, No.12105294, No.U1931201, No.12375107, No.12173039, the Department of Science and Technology of Sichuan Province, China No.24NSFSC2319, Project for Young Scientists in Basic Research of Chinese Academy of Sciences No.YSBR-061, 
and in Thailand from the NSRF via the Program Management Unit for Human Resources \& Institutional Development, Research and Innovation (B39G670013).

\noindent {\bf Author Contributions}\\
R.F. Xu and Kai Wang led the drafting of the manuscript and performed the data analysis for KM2A. Kai Wang also played a key role in cross-checking and conducted Fermi-LAT data analysis. S.Q. Xi and S.Z. Chen provided the analysis tools, good KM2A data, background map, and analysis  guidance. S.C. Hu for providing the WCDA spectrum points. R.Y. Liu helped with the theoretical interpretation and made significant contributions to revising the main text of the manuscript. Zhen Cao, the spokesperson of the LHAASO Collaboration and principal investigator of the LHAASO project. We would like to express our gratitude to R. Wang for her cross check for WCDA result and H.M. Zhang for assistance with the analysis. All other authors participated in data analysis, including detector calibration,data processing, event reconstruction, data quality checks, and various simulations, and provided comments on the manuscript.

\noindent {\bf Correspondent authors}\\
 R.F. Xu(xurenfeng@ihep.ac.cn), S.Z. Chen(chensz@ihep.ac.cn), Kai Wang(k-wang@smail.nju.edu.cn), R.Y. Liu(ryliu@nju.edu.cn), S.Q. Xi(xisq@ihep.ac.cn), S.C. Hu(hushicong@ihep.ac.cn)

\noindent {\bf Declaration of Interest}\\
The authors declare no competing interests.

\clearpage

\section{\bf Supplemental Information}\label{sec:sup_info}
\subsection{Introduction of the observatory LHAASO}

The LHAASO Experiment And Data Analysis: The Large High Altitude Air Shower Observatory (LHAASO) is a complex of arrays of Cosmic-ray(CR) and $\gamma$-ray detector located at Mt. Haizi ($4410 \rm{m} a.s.l.$, $29^{\circ}21^{'}27.56^{''} N$, $100 ^{\circ}08^{'}19.66^{''} E$) in Daocheng, Sichuan province, P.R. China. It consists of three sub-arrays, i.e., 1 km$^2$ array(KM2A), water Cherenkov detector array(WCDA), and wide field-of-view air 
Cherenkov telescope array(WFCTA). KM2A comprises an array of 5195 electromagnetic particle detectors(EDs) and an array of 1188 undersurface muon detectors(MDs) covering $\sim$ 1\,km$^2$. WCDA is constituted by 3120 detector units covering an area of about 78000 \,m$^2$. Combining with WCDA and KM2A, LHAASO can monitor and discovery the $\gamma$-ray sources located in northern sky (from about -20$^\circ$ to +80$^\circ$ in decl.), at energies between a few hundred GeV to above PeV with an unprecedented sensitivity, more 
details about detectors are presented in Ref\cite{ref_25}. Data used in this analysis were collected from  December 27, 2019, to July 31,2023 for KM2A and from March 2021 to October 2023 for WCDA. The data were reconstructed according to the method described in \cite{ref_13} for KM2A and in \cite{f17} for WCDA. We selected the $\gamma$-like events with $>$25 TeV KM2A based on the $\gamma$/p cut condition and applied a data quality cutting, detailed in published Crab Nebula analysis\cite{ref_13}; The KM2A events were divided into 12 logarithmic energy bins and into spatial bins with size of $0.1^{\circ} \times 0.1^{\circ}$. For each bins, the CR residual background were estimated by the direct integration method, which give the background number using the events in the same directions in horizontal coordinates but different arrival times. For the WCDA events, we selected that with the number of hits $> 300$\cite{f17}
and applied the $\gamma$-ray/background discrimination $\mathcal{P}_{incness}$ cut which is improved from \cite{f18} to refuse the most of CR events.  The WCDA events were divided into 3 group of the number of hits, in which the CR residual background are also given by the direct integration method. We performed a simulation used the CORSIKA and Geant4 package for above selected event sample. 

\subsection{Introduction of chance possibility}
To evaluate the association of the $\gamma$-ray source with the pulsar, we employ the method from \cite{1998ASPC..144...39M,ref_18} to estimate the probability of spatial coincidence by chance. The chance probability is related to the local pulsar density and the angular separation between the pulsar and the $\gamma$-ray source, and can be expressed as:
\begin{equation}
    P_{c}=1-e^{-r^2/r_0^2}
\end{equation}
where $r_0$ represents the characteristic angle between confusing sources, given by:
\begin{equation}
    r_0=[\pi\rho(\dot{E})]^{-1/2}
\end{equation}

The parameter $\rho(\dot{\rm{E}})$ represents the number density of pulsars. The relationship between $\dot{\rm{E}}$ of pulsars and VHE/UHE $\gamma$-ray sources is neither linear nor straightforward. however, most detected sources are related to energetic and close pulsars, such as those with $\dot{\rm{E}}/d^2 > 10^{34}\,\rm{erg\,s^{-1}\,pc^{-2}}$. In our case, we selected a $20\times5^{\circ}$ region to ensure sufficient sample size and minimize efficiency variations across the sky. If we ignore the $\dot{E}$ constraint, there are 11 samples that contain PSR~J1740+1000. If we apply a cut at $\dot{\rm{E}}/d^2> 10^{34}\, \rm{erg\,s^{-1}\,pc^{-2}}$, only one sample remains in the region. The pulsar density was calculated by dividing the area of the sky region and substituting the result into Equation (1) to derive the probability. Considering the possibility that the source lies within the tail region, approximately $60^{\circ}$ in range, the probability was further scaled by a factor of 1/6. The resulting probabilities in the two cases are 0.8\% and 0.07\%, respectively.

\subsection{Details for Fermi-LAT\rq s data analysis}

The Large Area Telescope (LAT) on board the Fermi satellite has continuously monitored the sky since 2008 and scans the entire sky every 3 hr \cite{ref_17}. This analysis uses Pass 8 data collected from August 4, 2008, to May 17, 2024, to study the GeV emission toward PSR~J1740+1000, where a Fermi source (4FGL J1740.5+1000) is located nearby. For the binned maximum likelihood analysis, all $\gamma$-ray photons within a $14^\circ \times 14^\circ$ region of interest centered on PSR~J1740+1000 are considered. The data analysis is performed using the publicly available software Fermitools (ver.2.0.8). The event types FRONT + BACK and the instrument response functions ($P8R3\_SOURCE\_V3$) are employed. To minimize contamination from 
$\gamma$-rays originating from the Earth's limb, a maximum zenith angle of $90^\circ$ is used. All sources listed in the fourth Fermi-LAT catalog are included, along with the diffuse Galactic interstellar emission (IEM, $gll\_iem\_v07.fits$) and isotropic emission ($iso\_P8R3\_SOURCE\_V3\_v1.txt$) in the background model. The spectral parameters of sources within $4^\circ$ of PSR~J1740+1000, as well as the Galactic and isotropic diffuse emission components, are all set free.

PSR~J1740+1000 is a $\gamma$-ray pulsar. The background source, 4FGL~J1740.5+1005, is associated with this pulsar, and the distance between them is only 0.1$^\circ$. To reduce contamination caused by the pulsar's own radiation, we have selected an energy range of 10-500 GeV for our analysis. The 4FGL source is subtracted as background from the point source analysis. The 4FGL~J1740.5+1005 is subtracted in the analysis as a point source. No clear extended emission has been detected, and the $95\%$ flux upper limits have been derived, assuming a point-like source template located at LHAASO-KM2A in the relevant energy band. The results are also shown in Fig. \ref{fig:SED}.
\begin{figure}[h]
    \centering
    \includegraphics[width=0.7\textwidth]{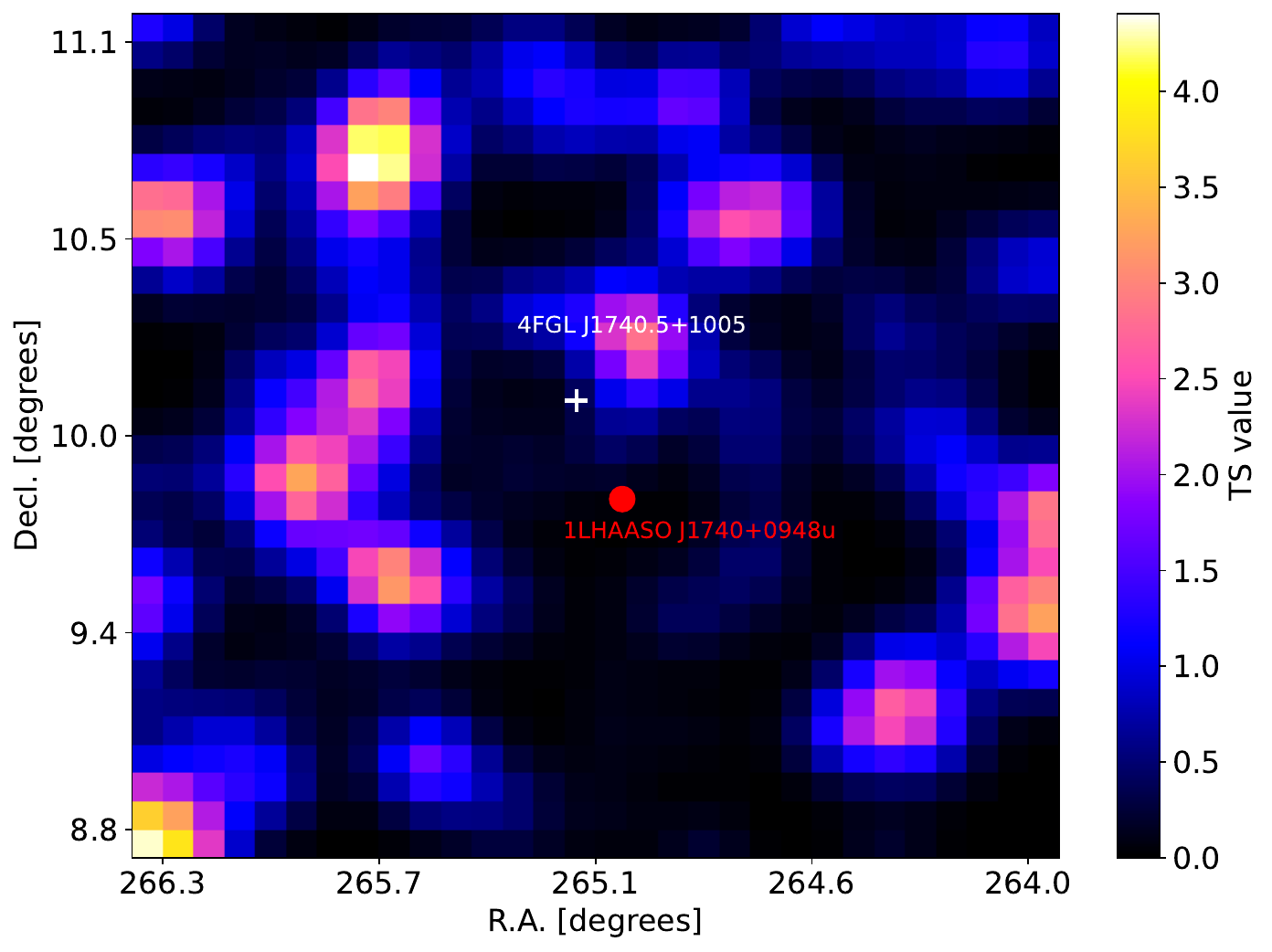}
    \caption{The residual TS map of the Fermi-LAT. 4FGL~J1740.5+1005 sources are shown with white crosses. The position of the 1LHAASO ~J1740+0948u is shown with a red dot.}
    \label{GeVTSmap}
\end{figure}

\subsection{Systematic error}

\begin{figure}[h]
	\begin{center}
		\includegraphics[width=0.72\textwidth]{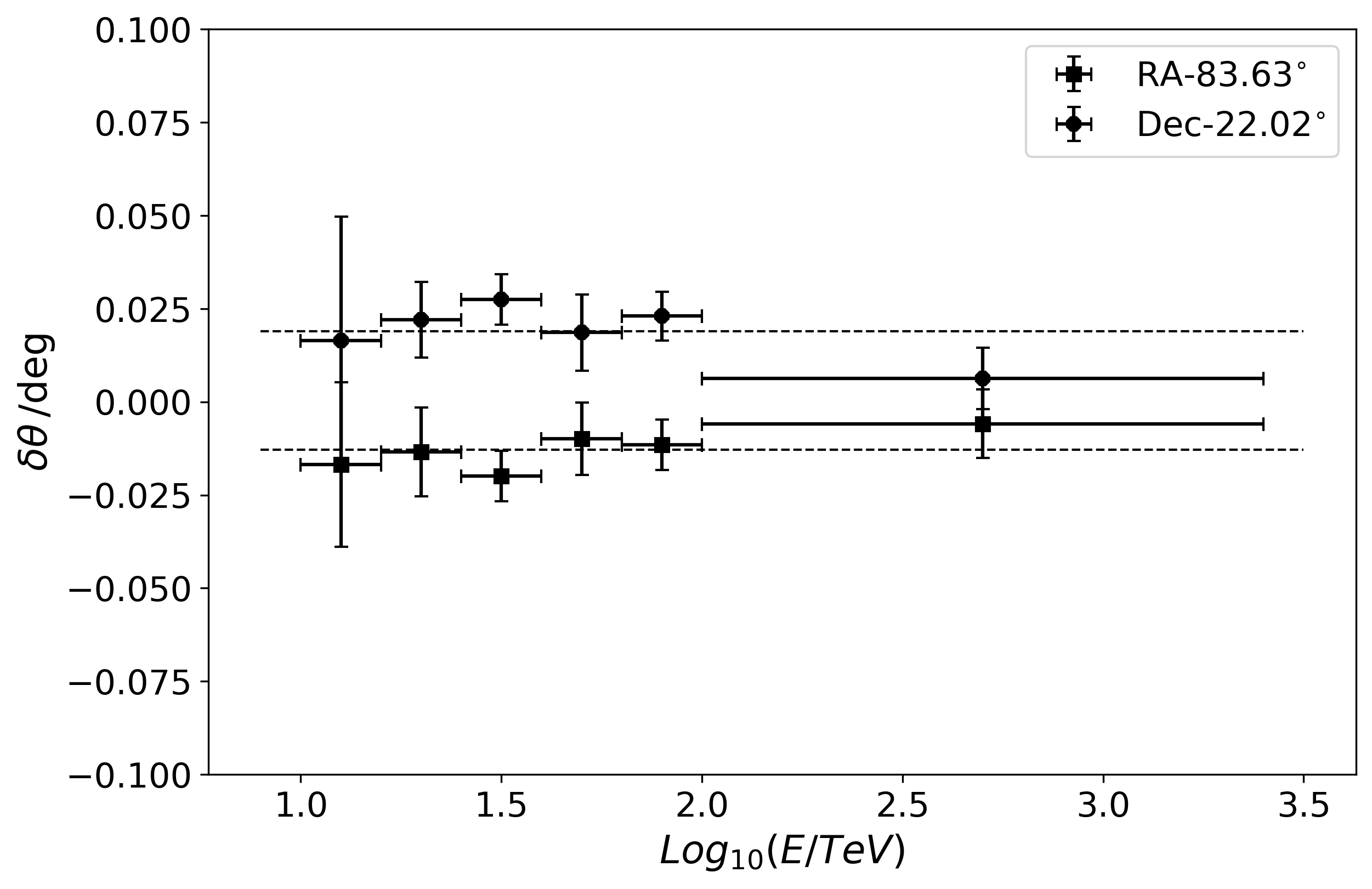}
    \end{center}
	\caption{In this study, we adopt the observed position deviation of Crab by LHAASO-KM2A as the local systematic error. } 
\label{fig:err}
\end{figure}

Systematic uncertainties: The Crab nebula was selected as the standard source for assessing the accuracy of the position fitting across different energy intervals. The comparison of the Crab's fitting position with its reference position, as illustrated in Figure \ref{fig:err}, revealed a negligible error of approximately $0.018^\circ$ in both RA and Dec, further corroborating the significant offset observed, this systematical error is consistent with the report value in 1LHAASO catalog\cite{ref_18}. Additionally, the systematic errors associated with the flux estimation were thoroughly addressed in previous work \cite{ref_13}, which reported errors of $0.08 \times 10^{-14}$ and $0.02$ for the normalization and spectral index, respectively. These systematic uncertainties were meticulously accounted for in our analysis, ensuring the reliability and robustness of the obtained results.


\subsection{Different epoch data analysis}
\begin{figure}[h]
    \centering
    \includegraphics[width=0.7\textwidth]{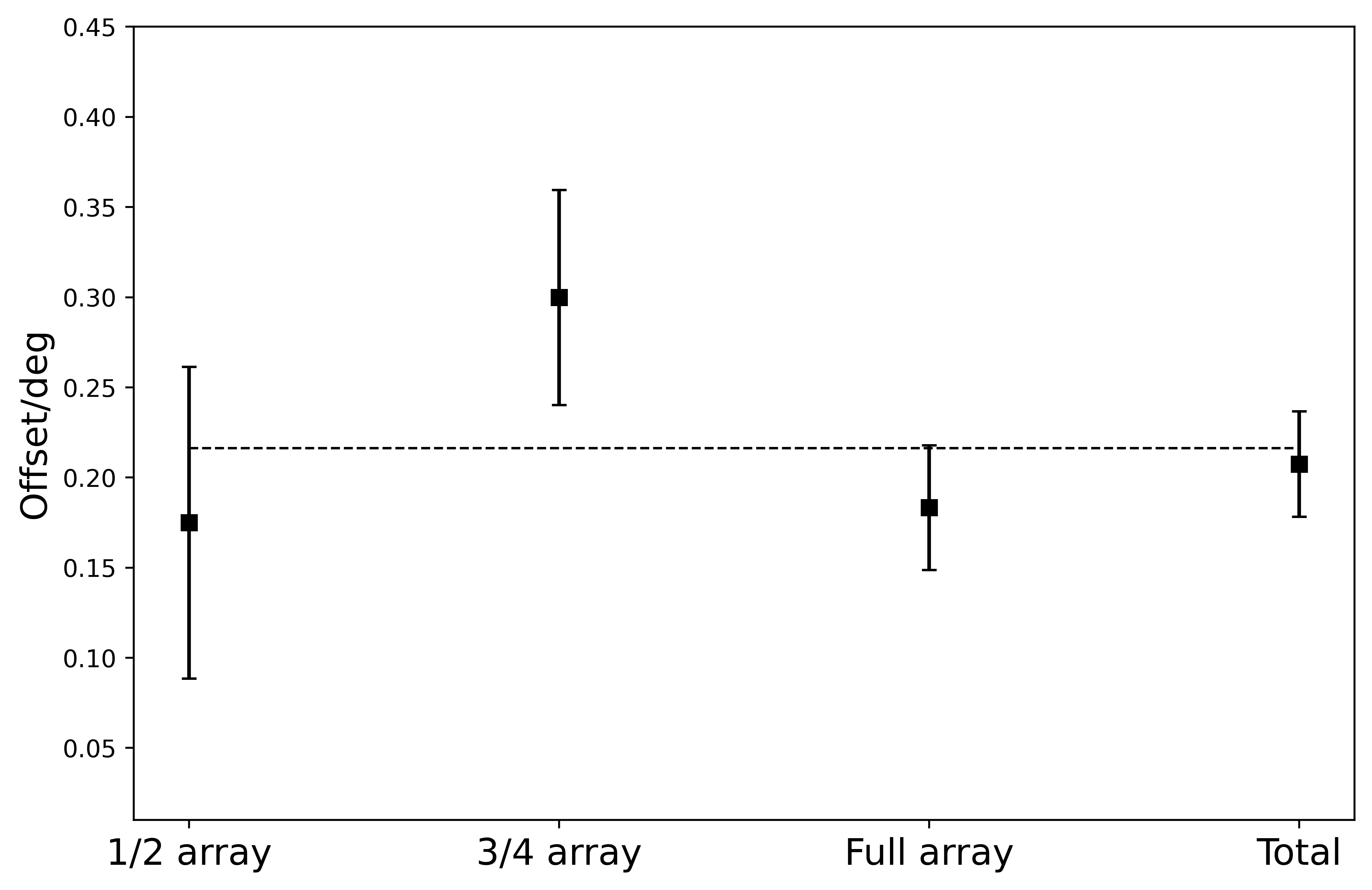}
    \caption{The left, middle, and right panels display the spectral, position, and positional offset fitting results using data from different array epochs.}
    \label{fig:diffarr}
\end{figure}

During different phases of the LHAASO project, the KM2A array expanded from half to three-quarters and eventually to full coverage. We applied different cuts and simulation data to select events, following the approach of previous work, and fit the data using the same pipeline. The fitting results, presented below, demonstrate that the source parameters remain consistent within the error margins across various epochs and array configurations. The offset values, consistently around 0.2°throughout different epochs, indicate the stability of the results(see Fig.\ref{fig:diffarr}).

\subsection{Gas distribution}
\begin{figure}[h]
    \centering
    \includegraphics[width=0.65\textwidth]{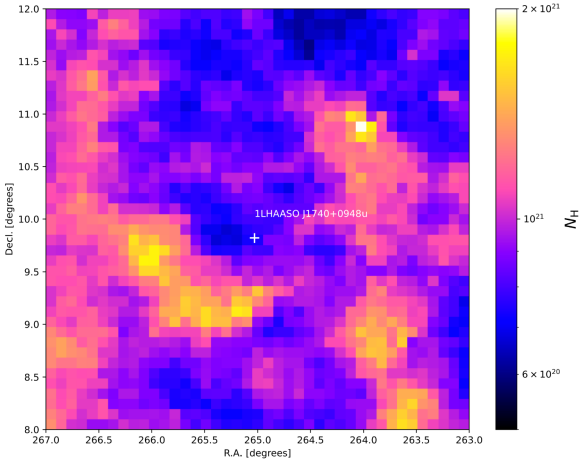}
    \caption{Column density map around PSR~J1740+1000 extracted from Planck dust data.}
    \label{fig:dust}
\end{figure}

The vibration rotation CO emission lines are commonly used tracers for molecular material. Since our target source is located in a high Galactic latitude region, most CO survey experiments do not provide coverage. We use the all-sky Planck data\cite{plank_dust1,plank_dust2} to trace the distribution of molecular clouds. To search for evidence that TeV emission originates from hadronic radiation, we should plot the CO distribution at the distance of the target source in this region. However, the lack of velocity information prevents us from accurately providing a profile at the current distance. Therefore, we present the dust column density distribution along the direction of the target source to simply compare the relationship between TeV emission and gas distribution. The column density $\rm{N_{H}}$ includes contributions from $\rm{N_{HI}}$ and $\rm{N_{H_2}}$, and can be expressed as $\rm{N_{H}}=\rm{N_{HI}}+\rm{N_{H_2}}$. $\rm{N_{HI}}=\frac{\tau_{353}}{\sigma_{353}}$, where $\sigma_{353}\approx \rm{10^{-26}\,cm^2\,H^{-1}}$ is used as an average value to estimate the approximate column density along the line of sight, and $\tau_{353}$ is the dust optical depth at 353\,GHz. ,where $\rm{N_{H_2}=2X_{CO}W_{CO}}$, where $\rm{W_{CO}=2\times 10^{20}\,cm^{-2}\,K^{-1}\,km^{-1}}$ is adopted as the mean CO-to-H2 mass conversion factor. The column density distribution calculated from this is shown in Fig.\ref{fig:dust}. In the region of the target source, the gas distribution is sparse, which is unfavorable for a hadronic origin of the TeV emission.
\end{document}